\documentclass[twocolumn,letter]{jpsj2}

\usepackage{bm}

\title{Nucleation of Vortex State in Ru-inclusion 
in Eutectic Ruthenium Oxide Sr$_2$RuO$_4$-Ru}

\author{Hirono \textsc{Kaneyasu}$^{1,2}$\thanks{Email: hirono@sci.u-hyogo.ac.jp}and Manfred \textsc{Sigrist}$^2$}
\inst{$^1$ Department of Material Science, University of Hyogo, Kamigori, Ako, Hyogo 678-1297, Japan \\
$^2$ Theoretische Physik, ETH-Z\"{u}rich, Z\"{u}rich CH-8093, Switzerland}

\date{\today}

\abst{Eutectic samples of Sr$_2$RuO$_4$-Ru possess many $\mu m $-sized Ru-metal inclusions. Based on a Ginzburg-Landau formulation we analyze the interplay of the chiral $p$-wave state of Sr$_2$RuO$_4$ ($T_{c} = 1.5 K $) and the s-wave state of Ru metal ($T_{c,Ru} = 0.5 K $) for an inclusion of cylindrical geometry. As a consequence of the mismatch of the order parameter phase, the occurrence of a spontaneous flux distribution appears for $ T < T_{c,Ru} $ which evolves into a ''Josephson vortex'' on the Sr$_2$RuO$_4$-Ru interface. At a sufficiently low temperature a depinning transition can occur whereby the vortex moves to the center of the cylinder.}

\kword{unconventional superconductivity, chiral p-wave, vortex}

\begin{document}

\maketitle
%

Sr$_2$RuO$_4$ is considered a textbook case of an unconventional superconductor. Over many years evidence has accumulated that its pairing symmetry is of chiral $p$-wave type, i.e. spin-triplet pairing with broken time reversal symmetry, generally denoted by the vector $ \vec{d} = \hat{z} (k_x \pm i k_y )$ ($p_x \pm i p_y $).\cite{Mackenzie-Modphys,Maeno-Nature,Maeno-Today}. The critical temperature is strongly sensitive to impurities and reaches $ T_c = 1.5 K $ in the cleanest samples. It came as surprise when Maeno and coworkers discovered that in eutectic samples of Sr$_2$RuO$_4$ with $\mu m $-sized Ru-metal inclusions the onset of inhomogeneous superconductivity could be found at a temperature roughly twice that of the bulk critical temperature\cite{Maeno-PRL}. Consequently, this superconducting phase was dubbed ''3-Kelvin phase'' (3K-phase). It was early speculated that the 3K-phase had filamentary nature and originates from nucleation of superconductivity at the interface between Ru-metal and Sr$_2$RuO$_4$\cite{Sigrist-JPSJ}. Ru-metal itself is a conventional superconductor with at $ T_{c,Ru} = 0.5 K $. It has not so far been possible to identify the microscopic mechanism which causes the increased transition temperature of the 3K-phase at the interface. 

Unlike in conventional superconductors where inhomogeneous nucleation of superconductivity leads through simple percolation eventually to a bulk superconducting phase, it was suggested that the unconventional nature of
Cooper pairing yields a more complex evolution from the 3K-phase to bulk superconductivity in Sr$_2$RuO$_4$ as temperature is lowered \cite{Sigrist-JPSJ,Kaneyasu}. Indeed there is even a symmetry breaking transition on the way to bulk superconductivity, since the nucleation of a $p$-wave pairing state on the Sr$_2$RuO$_4$-Ru interface leads to a time reversal conserving phase incompatible with the bulk superconducting phase. This second transition can actually be identified in experiment, e.g. in anomalies in the critical current \cite{Mao-PRL} and in quasiparticle tunneling \cite{Kawamura-JPSJ,Yaguchi-JPSJ,Yaguchi-AIP} as was shown in Ref.~\cite{Kaneyasu}. 

\begin{figure}
\begin{center}
\includegraphics[width=0.50\textwidth]{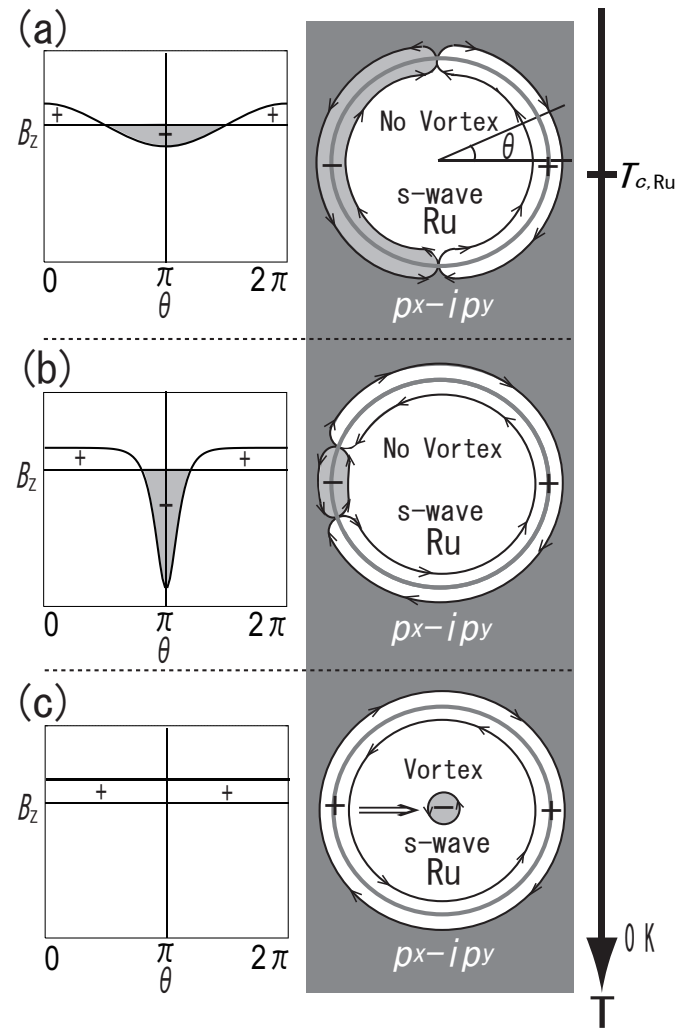}
\end{center}
\caption{Flux distribution and vortex state: (a) immediately below $ T_{c,Ru} $ a sinusoidal flux pattern appears with two nodes diametral on cylindrical interface; (b) at lower temperature gradually a more concentrated flux line emerges on the interface; (a) and (b) correspond to a state of  $ \psi $ without phase winding; (c) below a critical temperature the flux line depins from the interface and forms a vortex at the center of the cylinder, a state with phase winding of $ \psi $.} 
\label{Fig-Vortex}
\end{figure}

So far the situation for these eutectic sample at low temperature, where also the Ru-inclusions become superconducting on their own, has not drawn much attention. In this study we would like to address one particular aspect of this situation which is connected with the fact that actually the two superconducting phases, $s$-wave for Ru and chiral $p$-wave for Sr$_2$RuO$_4$, meeting at interface are not phase compatible, a feature we will explain below. We would like to demonstrate here that this could lead to the 
creation of intriguing magnetic flux distributions which may undergo a transition between two distinct states as temperature is lowered. 

We illustrate this idea using the convenient geometry of a cylindrically shaped Ru-metal inclusion in an infinitely large matrix of Sr$_2$RuO$_4$, which is most suitable for the chiral $p$-wave phase. The axis of the cylinder lies along the $z$-axis and the radius $R$ shall be of order of several $ \mu m $, a dimension we assume to be large compared coherence lengths of both superconductors. When Ru becomes superconducting the $p$-wave superconducting state of Sr$_2$RuO$_4$ is already solidly established and will be considered as rigid here. Looking now at the superconducting phase of Ru we may model it by the following
Ginzburg-Landau (GL) free energy.  We choose here cylindrical coordinates $(r,\theta, z)$:
\begin{equation} \begin{array}{ll}
F  = & \displaystyle \int_{Ru} dr \; d\theta \; r \left[ a |\psi|^2 + b | \psi|^4 + K | \vec{D} \psi|^2 + \frac{\vec{B}^2}{8 \pi} \right] \\ & \displaystyle + \int_{int} d\theta \; R \; \tilde{K} \{ \hat{z} \cdot (\vec{n} \times \vec{\eta})^* \psi + c.c. \}
\end{array}
\end{equation}
where $ F $ denotes the free energy per unit length along the $z$-axis with real coefficients $a = a'(T-T_{c,Ru}) $, $b$ and $ K $ and
the covariant derivative $ \vec{D} = \vec{\nabla} - i 2e \vec{A} /hc $ (neglecting $z$-dependence). The first part is the standard GL free energy for the order parameter  $ \psi $ of the conventional superconducting phase within the Ru-inclusion. The second term is for our purpose most essential as it represents the coupling between the order parameters for the two materials with the coupling constant $ \tilde{K}$. At this level it
describes the lowest-order Josephson coupling \cite{Geshkenbein,Sigrist-RMP}. 
For the chiral $p$-wave state we define $ \vec{\eta} = (\eta_x, \eta_y,0) = \eta_p (1, \pm i,0) $ 
with  a complex constant amplitude $ \eta_p $ and $ \vec{n} $ as the normal unit vector of the interface. Note that for this Josephson coupling spin-orbit coupling is important as it provides the connection between the spin-singlet and spin-triplet configurations of the Ru $s$-wave and the Sr$_2$RuO$_4$ $p$-wave phase, respectively. The structure of the coupling incorporates the selection rule of conserved total angular momentum $ J_n $ of the Cooper pairs along the interface normal ($ \vec{J} \cdot \vec{n}= J_n $ conserved) \cite{Geshkenbein,Sigrist-RMP}.

The interface term imposes boundary conditions for $ \psi $ at $ r = R $,
\begin{equation}
\left. K \vec{n} \cdot \vec{D} \psi \right|_{r=R} = \left. \tilde{K} \hat{z} \cdot (\vec{n} \times \vec{\eta} ) \right|_{r = R} \; .
\end{equation}
Taking the chiral $p$-wave component as rigid and $ \vec{n} = (\cos \theta, \sin \theta,0) $ we may write for the term on the right hand side,
\begin{equation}
\hat{z} \cdot ( \vec{n} \times \vec{\eta} ) = -i \eta_p e^{i \theta} = | \eta_p | e^{i (\theta - \alpha)} 
\end{equation}
choosing among the two chiral phases, the $ p_x - i p_y $ state  (with a gauge transformation we may choose $ \alpha =0 $ which is equivalent to the rotation of the coordinate frame around the $z$-axis by the angle $ - \alpha $). The resulting boundary condition implies the competition between two low energy configurations for $ \psi $. One option is a
state without angular dependence of $ \psi$ which  ignores the coupling to the phase imposed by the chiral $p$-wave state through the interface. 
This state is stabilized at the expense of interface energy. 
The other is to pick up the phase winding of the chiral $p$-wave state with the disadvantage of having to introduce an energetically costly singular line (vortex) which for symmetry reason would lie on the axis of the cylinder ($r=0$). If the coupling at the interface is weak, it is the former phase which nucleates at $ T=T_{c,Ru} $
and is stable for some temperature range. At lower temperature the other phase becomes competitive and may eventually win. 

In order to obtain a qualitative understanding of the evolution of these states we focus now on
the interface assuming the order parameter $ \psi $ to be finite and rigid in its modulus, such that we may restrict to discussion of a variable order parameter phase,
\begin{equation}
\psi (r=R,\theta) = | \psi (R)| e^{i \phi(\theta)} 
\end{equation}
at $ r = R $. Then we can derive in a standard way the effective free energy functional  for the phase $ \phi (\theta) $ at the interface which 
has the form
\begin{equation} 
F_{int} =  f_0 \int_{0}^{2 \pi} d\theta \;   \left[ \frac{1}{2}\left( \frac{d \phi}{d \theta} \right)^2 - \Lambda^2 \cos(\phi - \theta) \right]
\label{fpth}
\end{equation}
with the energy scale 
\begin{equation}
f_0 =  \left( \frac{\Phi_0}{2 \pi \sqrt{d_{eff} R}} \right)^2 
\end{equation}
and $  \Lambda^2 f_0 = 2 \tilde{K} | \psi | | \eta_p| $ where $ d_{eff} = \lambda_s + \lambda_p + d $ is the effective width of the interface for magnetic fields including the two London penetration depths, $ \lambda_{s,p} $ and the width of the interface region, $d$ \cite{Tinkham}. The temperature dependence of $ \Lambda^2 $ is governed by the order parameter $ | \psi (T)| $ and the London penetration depth $ \lambda_s(T) $. We consider the other quantities such as $ | \eta_p | $ and $ \lambda_p $ as basically independent of
temperature. Then the evolution as temperature is lowered corresponds to a monotonic increase of $ \Lambda^2 $  from zero at $ T = T_{c,Ru} $. We introduce here also the magnetic flux quantum $ \Phi_0 = hc / 2e $ of a superconductor.  

Let us now discuss the behavior of the phase $ \phi (\theta) $ at the interface in our simplified model. 
The variation of Eq.(\ref{fpth}) with respect to $ \phi(\theta) $ leads to 
\begin{equation}
\frac{\partial^2 \phi}{\partial \theta^2} = \Lambda^2 \sin(\phi(\theta) - \theta) \; .
\label{varp}
\end{equation}
Note that the derivative of $ \phi(\theta) $ with respect to $ \theta $ corresponds to the local magnetic field in $z$-direction,
\begin{equation}
B (\theta) =\sqrt{f_0} \frac{\partial \phi (\theta) }{\partial \theta} \; ,
\end{equation}
like in standard Josephson junctions \cite{Tinkham}.
We turn now the solution of the variational equation.  Within the given boundary condition that $ \phi(\theta) = \phi(\theta + 2 \pi) $ without any phase winding for the order parameter $ \psi $. We consider approximative solutions for two limiting cases,
\begin{equation}
\phi(\theta) \approx \left\{ \begin{array}{cl} \Lambda^2 \sin \theta & {\rm for} \; \Lambda \ll 1  \; , \\ & \\
\theta - 4 \arctan[e^{\Lambda(\theta - \pi)}] & {\rm for} \; \Lambda \gg 1  \; . \end{array} \right.
\end{equation}
which yields to the magnetic field distribution
\begin{equation}
B(\theta) \approx \sqrt{f_0} \left\{ \begin{array}{cl} \Lambda^2 \cos \theta & {\rm for} \; \Lambda \ll 1  \; ,\\ & \\ 
\displaystyle 1 - \frac{2 \Lambda}{\cosh^2 [ \Lambda(\theta - \pi)]} & {\rm for} \; \Lambda \gg 1 \; . \end{array} \right. 
\end{equation}
These solutions show a magnetic flux distribution located around the interface. The net flux vanishes due to flux quantization in the superconductor, since there is no phase winding of the order parameter of Sr$_2$RuO$_4$. With decreasing temperature the flux distribution changes from a soft sinusoidal form to a more and more uneven structure where one magnetic field direction becomes spatially more concentrated evolving into a Josephson vortex, and the magnetic flux in opposite direction is more spread (Fig.\ref{Fig-Vortex} a,b). 
Also the local magnetic fields grow with shrinking temperature. 
Note that in our model geometry there is a rotational degeneracy for this flux distribution.  

\begin{figure}
\begin{center}
\includegraphics[width=0.50\textwidth]{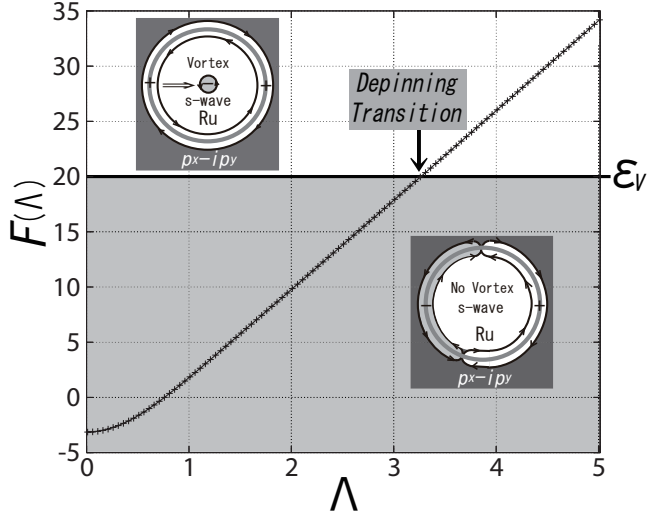}
\end{center}
\caption{Free energy balance for the interface state: for small values of $ \Lambda $ (''high'' temperature) the state of $\psi $ without phase winding (flux distribution restricted to the interface) is realized; for large enough $ \Lambda$ the vortex state with phase winding of $ \psi $ is stabilized. The transition can be viewed as a spontaneous flux line depinning transition.} 
\label{Fig-energy}
\end{figure} 

While the state with this topology is stable for some range of temperature below $ T_{c,Ru} $, the alternative state with an order parameter $ \psi (r,\theta) = | \psi(r) | e^{i \theta} $ possessing a single
vortex at the center of the cylinder becomes energetically more competitive at lower temperature. 
The energy of this vortex state can be estimated as
\begin{equation}
F_{v} = 2 \pi f_0 \left(\frac{1}{2}  - \Lambda^2 \right) + \epsilon_v \; ,
\end{equation}
where the first term is obtained from Eq.(\ref{fpth}) by setting $ \phi(\theta) = \theta $. The second term
is the line energy per unit length of the vortex which we approximate by
\begin{equation}
\epsilon_v \approx \left( \frac{\Phi_0}{4 \pi \lambda} \right)^2 \ln \kappa + \epsilon_c
\end{equation}
with $ \kappa = \lambda / \xi $ and $ \epsilon_c $ the core energy which is a fraction of the first term constituting the magnetic contribution to the line energy. 

In Fig.\ref{Fig-energy} we show the energy $ F = F_{int} - 2 \pi f_0 (1/2 - \Lambda^2) $ obtained by numerical solutions of the Sine-Gordon equation in Eq.(\ref{varp}). For $ \Lambda > 1 $ the energy follows an
essential linear behavior: $ F/f_0 \approx -7.2 + 8.2 \Lambda $.  
The transition between the two states is determined by $ F \approx \epsilon_v $, i.e. the condition when the line energy of the vortex at the center of the Ru-inclusion and the energy to carry the flux distribution on the interface are equal. Neglecting the core energy $ \epsilon_c $,
taking $ \ln \kappa \sim 1 $ and $ d_{eff} \sim 2 \lambda $, we obtain the rough criterion,
\begin{equation}
\frac{F (\Lambda)}{f_0}  \sim \frac{R}{2 \lambda} \; .
\end{equation}
Assuming a low-temperature value of $ \lambda \sim 0.2 \mu m $ and $ R \sim  10 \mu m $ we estimate from Fig.\ref{Fig-energy} that the criterion is satisfied for $ \Lambda \approx 3.2 $. It is obvious that with increasing $ R$ the critical value of $ \Lambda $ for the transition grows. Consequently, for large Ru-inclusions the transition between the two states is less likely to occur than for smaller inclusions.
The state after the transition corresponds to a single vortex at the center of the Ru-inclusion with a compensating magnetic flux on the interface to yield an overall vanishing flux (see Fig.\ref{Fig-Vortex}c). 

In summary our study shows that the coupling of the intrinsic superconducting order parameter of a Ru-inclusion in the eutectic ruthenate samples is subject to a frustrating coupling with the $p$-wave order parameter of the surrounding Sr$_2$RuO$_4$ inducing states with a spontaneous magnetic flux distribution. There are two competing states which both induce a spontaneous flux pattern: the interface vortex state and the Ru-center vortex state, which are distinguished by the topology of the order parameter $ \psi $ (phase winding) of the Ru-superconductor. 
We show here that there is a clear sequence, how these states would appear as temperature is lowered. 
First the interface vortex phase appears below $ T_{c,Ru} $and only at lower temperature a transition to the center vortex state occurs. This transition can be viewed as a depinning transition of the interface vortex which moves to the center of the cylinder and leaves behind a compensating counter flux uniformly spread over the interface. This suggests also that
the additional pinning effects, through interface defects or special conditions due to varying interface curvature, would play an essential role too in defining the transition point. Such issues are certainly important  for inclusions of rather irregular shape. Even under these condition transitions between different
stable flux configurations are possible, although the situation is considerably more complex and beyond the scope of our study. We believe that these effects should be accessible to studies using scanning Hall or SQUID probes.

We are grateful to N. Hayashi, K. Makoshi, Y. Maeno and H. Yaguchi for helpful discussions. The numerical calculations were carried out on SX8 at YITP, Kyoto University. This study was supported by Suzuki Foundation, the Japan Securities Scholarship Foundation and the Swiss Nationalfonds through the NCCR MaNEP and the Center for Theoretical Studies of ETH Zurich.

\end{document}